\documentclass{aa}
\usepackage{graphicx}
\begin{document}


  \title{Semi-classical theory of collisional depolarization of spectral lines by atomic hydrogen. I: application to $p$ states of neutral atoms.}

   \author{M. Derouich
          \inst{1},
           S. Sahal-Br\'echot
          \inst{1},
          P. S. Barklem
          \inst{2},
           and
          B. J. O'Mara
          \inst{\dag 3}
          }
\titlerunning{Semi-classical theory of collisional depolarization .}
\authorrunning{M. Derouich et al}
   \institute{Observatoire de Paris-Meudon, LERMA FRE CNRS 2460, 5, Place Jules Janssen, F-92195 Meudon Cedex, France. 
          \and
Department of Astronomy and Space Physics, Uppsala University, Box 515, S 
751 20 Uppsala, Sweden
           \and
Department of Physics, The University of Queensland, St Lucia 4072, Australia\\
   \email{Moncef.Derouich@obspm.fr}
             }
   \date{Received  2002 / accepted 2003}

\abstract{
The present paper extends the  method of Anstee, Barklem and O'Mara (Anstee \cite{Anstee2}; Anstee \& O'Mara \cite{Anstee1}, \cite{Anstee3}; Anstee, O'Mara \& Ross \cite{Anstee4}; Barklem \cite{Barklem3}; Barklem \& O'Mara \cite{Barklem1}; Barklem, O'Mara \& Ross \cite{Barklem2}), 
developed during the 1990's  for collisional line broadening by atomic 
hydrogen, to  the depolarization of spectral
 lines  of neutral atoms by collisions with atomic hydrogen. In 
the present paper, we will limit the calculations to $p$ ($l=1$) atomic levels. The depolarization cross sections and depolarization rates are computed. In 
Table \ref{cross}  cross sections as functions of the relative velocity and 
effective quantum number are given, allowing for the computation for any $p$ atomic 
level. Our results are compared to quantum chemistry  calculations  where possible. The sensitivity
 of depolarization cross sections to regions of the potential is examined.
We conclude that the accuracy obtained with our method ($<$ 20 \% for the depolarization rates) is 
promising for its extension to higher $l$-values for the interpretation of the ``second solar spectrum''. This will be the object of further papers.  

}
\maketitle


\section{Introduction} \label{sec1}
 
Observations of linear polarization at the limb of the Sun produce  rich 
structures. This polarization spectrum  (the so-called ``second solar spectrum'')  is an entirely new spectrum to be explored
and interpreted (Stenfo \& Keller \cite{Stenflo}).
This polarization is due to  scattering of the incident anisotropic  solar 
radiation field. The radiative anisotropy is due to the structure of the
semi-infinite or layer atmosphere in the vicinity of the surface, and is 
revealed by the center-to-limb darkening of the solar brightness.

In order to interpret quantitatively the Stokes parameters of the observed 
lines, a non-optically-thin and non-LTE model in the presence of a magnetic 
field has to be used. In fact, collision induced transitions between Zeeman sublevels can play a major 
role in depolarizing the levels (and thus the lines). Depolarizing	collisions of the 
radiating and absorbing atoms with neutral hydrogen atoms of the medium have 
to be included in the statistical equilibrium equations. Very few depolarization
 rates  have currently been computed, they have been obtained through sophisticated quantum
 chemistry methods which are accurate but cumbersome. It would be 
useful to develop methods capable of giving reasonable 
results for many levels rapidly. This is the object of the present work.

  During the 1990's  Anstee, Barklem and O'Mara (Anstee \cite{Anstee2}; Anstee \& O'Mara \cite{Anstee1}, \cite{Anstee3}; Anstee et al. \cite{Anstee4}; Barklem \cite{Barklem3}; Barklem \& O'Mara \cite{Barklem1}; Barklem et al. \cite{Barklem2}) developed a new semi-classical theory, in order to obtain spectral line widths due to collisions with atomic hydrogen that were better than those obtained by the 
Van der Waals approximation. In this theory, which will be referred as ABO in 
the following, the scattering $S$ matrix is obtained by solving semi-classical
coupled differential equations.    
The hydrogen-atom interaction potential is derived from time-independent second-order 
perturbation 
theory without exchange, allowing  the Lindholm-Foley average 
over $m$ states (Brueckner \cite{Brueckner}; O'Mara \cite{O'Mara}) to be removed. 
Whereas the Van der Waals potential underestimates the broadening by typically a factor two or more, the ABO method gives  rather good agreement ($<$ 20 \% for the spectral line widths) 
with the results obtained from quantum chemistry calculations. In addition, these
authors used their ABO method by applying it to a selection of strong solar 
lines. The derived abundances are consistent with meteoritic ones.

In the present paper, we extend the ABO theory to the calculation of collisional depolarization rates which enter the statistical equilibrium equations for the polarized atomic density matrix. This paper is focused  on the method and is applied to $p$ states of neutral atoms. We will
present and compare our results with those recently obtained by 
quantum chemistry methods. We also compare the results to the depolarization 
rates calculated with the Van der Waals interaction. A discussion of 
our results is given, studying the region of the potential curves which 
play the major role in the variation of the depolarization cross sections. 
 In conclusion, we show that our method, 
owing to the simplicity of the processes taken into account, is a useful 
alternative to the accurate but time consuming quantum chemistry method for obtaining 
most of the depolarization rates to better than 20 \% accuracy. It should now be
possible to rapidly obtain the large amount of data needed for the interpretation
of the second solar spectrum, in particular for the heavy atoms that are inaccessible to  present quantum chemistry calculations.
An extension to $d$ and $f$ states is in progress. An extension to ions is
in progress and this will be the subject of further papers.
\section{Formulation of the problem } \label{sec2}
\subsection{The statistical equilibrium equations (SEE) for polarization studies}
The calculation of the Stokes parameters of a spectral line observed at the
solar  limb is a non-LTE (Local Thermodynamical Equlibrium) problem. It 
needs 
the solution of the coupling between the polarized radiative transfer and the statistical equilibrium equations (SEE) for the populations and coherences of the polarized atomic density matrix. In fact the atomic levels are only ``aligned'', which means that the observed lines are globally linearly polarized. This linear 
polarization is due to anisotropic scattering of the incident solar radiation 
field which contributes to excitation of the atomic levels, the anisotropy
being due to limb darkening, even  at zero altitude above the limb.

 Sahal-Br\'echot (\cite{Sahal2}) developed the SEE for an ``aligned'' multilevel atom, i.e.
 for the Zeeman sublevels $|\alpha J  M_J\rangle$. She took into account excitation 
by the anisotropic solar radiation field, deexcitation by spontaneous and induced emission, and excitation, deexcitation, polarization transfer  and depolarization by isotropic collisions of the particles of the medium. She included the effect of a ``strong'' magnetic field (strong case of the Hanle effect) and thus coherences between Zeeman sublevels could be ignored. Only the populations of the Zeeman sublevels were  taken into account.

 She wrote these equations both in the dyadic basis $|\alpha J M_J\rangle \langle  \alpha' J' M'_J|$ built on the standard basis of the atomic states $|\alpha J M_J\rangle $ and in the irreducible tensorial operator basis $^{\alpha J \alpha' J'}T_q^k$. The expansion of the $^{\alpha J \alpha' J'}T_q^k$ over the dyadic basis is given by
Fano \& Racah (\cite{Fano1}) (cf. also Messiah \cite{Messiah}; Fano \cite{Fano2}):\\
$^{\alpha J \alpha' J'}T_q^k = \sum_{M_J M'_J} (-1)^{J'-M'_J}  \nonumber$ \\
\begin{eqnarray} \label{eq1}
 \quad \quad \quad \times  |\alpha J M_J \rangle  \langle \alpha' J' M'_J| 
\langle J J' M_J -M'_J|k -q \rangle,
\end{eqnarray}
where $\langle J J' M_J -M'_J|k -q \rangle$ is a Clebsch-Gordan coefficient.
She demonstrated the interest and power of this second basis, which is especially well adapted to problems with spherical and/or cylindrical symmetry processes. The
equations are simplified, and the physics is extracted and highlighted. In particular, the odd $k$-terms can be eliminated from the SEE when the incident  radiation field is anisotropic but unpolarized. In fact the natural radiation field transports $k=0$ (population) and $k=2$ (alignment) terms only. Orientation  terms ($k=1$)  vanish in the present astrophysical problem.

 Sahal-Br\'echot (\cite{Sahal2}) gave the general expressions for the various transition probabilities due to isotropic and anisotropic radiation and for isotropic
collisions, i.e. expressions for population, orientation and alignment transfer between the levels, creation and destruction of population, orientation and  alignment (depolarization rates in particular). At the same time, Omont (\cite{Omont2}) developed a general theory of relaxation for a two-level atom. In addition he introduced coherences.

In fact, for isotropic processes, the coupling terms in the SEE implying transfer of coherence $q \to q'$ are zero,  and due to the isotropy of the collisions with hydrogen, the depolarization rates are $q$-independent. This is one of the advantages of the irreducible tensorial operator basis. Therefore, as we will consider  isotropic 
collisions only in the following, it will be sufficient to calculate collisional cross sections between the sublevels, as for the no-coherence case (Sahal-Br\'echot \cite{Sahal2}), even if coherences play a role in the SEE. We recall that   coherences can appear when the magnetic field is weak and not directed  along the 
solar radius (Hanle effect). We also notice that  in the present paper we are interested in hydrogen-atom collisions and not in electron or proton-ion collisions as  in Sahal-Br\'echot (\cite{Sahal2}). After integration of the cross sections over a Maxwell distribution of velocities and multiplying by the local hydrogen density, the  depolarization rates obtained (and eventually polarization transfer rates between levels) can be used in the  statistical equilibrium equations.

 We  focus on the collisional depolarization rates in the present paper. We will use both the dyadic  and the irreducible tensorial operator bases. The latter one is in fact more convenient.

 An important preliminary point to notice is the fact that the radiating atom can have a nuclear spin and thus  hyperfine structure. The importance of the effect of the hyperfine structure on the Stokes parameters of the  line studied was discussed for the example of the hydrogen coronal line Ly$\alpha$  by Bommier \& Sahal-Br\'echot (\cite{Bommier2}). They showed that the SEE must be solved for the hyperfine levels when the inverse of the lifetime  of the upper level is smaller than the hyperfine splitting, i.e. the hyperfine levels are separated.
\begin{table}
\begin{center}
\begin{tabular}{|l|c|r|}
\hline
F & $\Delta E_{HFS}(^2P_{3/2})$ & $\Delta E_{HFS}(^2S_{1/2})$ \\
\hline
3 & $58.44$ & - \\
\hline
2 & $34.32$ & $1772$ \\
\hline
1 & $15.86$ & - \\
\hline
\end{tabular}
\end{center}
\caption{Values of the energy separation between the  hyperfine levels $F$ and $F-1$ for: Na in the ground state, $3s \;^2S_{1/2}$ (Ackermann \cite{Ackermann}), and in
the excited state $3p\; ^2P_{3/2}$ (Gangrsky et al \cite{Gangrsky}). All energies are in  MHz.}
\label{NaHFS}
\end{table} 
This arises from the Heisenberg uncertainty principle. On the contrary, when the  hyperfine splitting is small compared to the inverse of the lifetime, the hyperfine levels are mixed and degenerate, the hyperfine structure can be ignored and the SEE can be solved for the $| \alpha J \rangle$ fine structure levels. This was the case 
for the  hydrogen coronal line where the $F$ levels are degenerate and the $J$ levels are well separated.

In the  second solar spectrum, where metallic lines are observed,  the inverse of the lifetime (9.7 MHz for the Na D lines) (Gaupp et al \cite{Gaupp}) is often small compared to the 
hyperfine splitting (if it exists) (cf. Table \ref{NaHFS}). Then the SEE must be solved for the $| 
\alpha J F \rangle$ hyperfine levels (Bommier \& Molodij \cite{Bommier3}; Kerkeni \& Bommier \cite{Kerkeni3}) and  collisional rates between the hyperfine sublevels must be calculated. 
\subsection{Dyadic basis for the depolarization rates}
The depolarization arises from isotropic collisions between the Zeeman sublevels of a  level $| \alpha J \rangle $: the collisions reestablish thermodynamical equlibrium between the sublevels and thus atomic polarization vanishes. We have to calculate (Sahal-Br\'echot \cite{Sahal2}) for each sublevel $| \alpha J M_J \rangle $ (or $| \alpha J F M_F \rangle $ if there is an hyperfine structure):
\begin{eqnarray} \label{eq2} 
D (\alpha J  M_J, T) = \sum_{M'_J \ne M_J} \zeta (\alpha J  M_J \to \alpha J  M'_J, T)
\end{eqnarray}
where $\displaystyle \zeta (\alpha J  M_J \to \alpha J  M'_J, T)$ is the collisional transition rate between the sublevels $|\alpha J M_J \rangle \to |\alpha J  M'_J \rangle$. It 
can be written as a function of the local temperature $T$ and the hydrogen 
perturber local density $n_H$:
\begin{eqnarray} \label{eq3}
\zeta(\alpha J M_J \to \alpha J M'_J, T) = n_H &&\int_{0}^{\infty} \sigma (\alpha J M_J  \to \alpha J M'_J, v) \nonumber \\
&&  \times v f(v) dv,
\end{eqnarray}
$f(v)$ being the distribution function for  the atom-perturber relative velocity $v$, and     $\displaystyle \sigma (\alpha J M_J  \to \alpha J M'_J, v)$ is the cross section. For the Maxwellian distribution,
\begin{eqnarray} \label{eq4}
f(v)=\sqrt{\frac{2}{\pi}} (\frac{\mu}{kT})^{\frac{3}{2}} v^2 \textrm{exp}(-\frac{\mu v^2}{2kT}),
\end{eqnarray}
where $\mu$ is the reduced mass of the system.

The relaxation term for the density matrix (diagonal term of the system of 
SEE) due to collisions is given by:   
\begin{eqnarray} \label{eq5} 
\big(\frac {d ^{\alpha J}\rho_{M_JM_J}}{dt})_{coll}& = &- D (\alpha J  M_J , T) \quad ^{\alpha J}\rho_{M_JM_J}  \\
& & -   \sum_{\scriptstyle \alpha' J' \ne \alpha J} \sum_{\scriptstyle M'_J} \zeta (\alpha J  M_J \to \alpha J'  M'_J, T) \nonumber \\
&& \times  ^{\alpha J, \alpha' J'}\rho_{M_JM'_J} \nonumber 
\end{eqnarray}
The terms added to the depolarization term $ \displaystyle D (\alpha J  M_J, T)$ are the so-called quenching terms.
 
 The dyadic basis is not suitable because the depolarization terms depend on the chosen dyadic basis. The $\displaystyle ^{\alpha J, \alpha' J'}T_q^k$ basis is well adapted because the depolarization terms do not depend on the  basis and have a physical interpretation.
\subsection{Basis of irreducible tensorial operators}
In this basis, the SEE relaxation term  in the non-coherence case ($q=0$) (Sahal-Br\'echot \cite{Sahal2}; Omont \cite{Omont2}) becomes 
\begin{eqnarray} \label{eq6}
\big(\frac{d ^{\alpha J}\rho_0^{k}}{dt})_{coll}=- D^k(\alpha  J, T)^{\alpha J \alpha J}\rho_0^k + \textrm{quenching term,}   
\end{eqnarray} 
where $\displaystyle ^{\alpha J}\rho_0^{k} $ is the multipole $k$-moment of the density matrix for the level $|\alpha J \rangle $. For isotropic collisions $\displaystyle D^k(\alpha  J, T)$ is characterized by $(2 J+1)$ constants ($0 \le k \le 2 J$) and
$D^0(\alpha  J, T)$ is called the destruction rate of population, which is zero for the no-quenching approximation. $D^1(\alpha J, T)$ is the destruction rate of orientation (circular atomic polarization). $D^2(\alpha  J, T)$ is the destruction rate of alignment which is of interest 
in astrophysics (linear polarization), and  the depolarization rate  $\displaystyle D^k(\alpha  J, T)$ is given by:
\begin{eqnarray} \label{eq7}
D^k(\alpha  J, T)= n_H \int_{0}^{\infty} \sigma^k(\alpha J, v) 
v f(v) dv  
\end{eqnarray}
Using our present notation, formula (20) in Sahal-Br\'echot (\cite{Sahal2}) with $k = k'$ can be 
written as:
\begin{eqnarray} \label{eq8}
\sigma^k(\alpha  J, v)& = & (2k+1) \sum_{\scriptstyle M'_J} \sum_{\scriptstyle M_J \ne M'_J}  (-1)^{2J-2M_J} \nonumber \\
&& \times \left(
\begin{array}{ccc} 
J & k &  J  \\
-M_J &  0 & M_J  
\end{array}
\right)^2  \sigma (\alpha J  M_J \to \alpha J  M'_J, v)   \nonumber \\
&& - (2k+1) \sum_{\scriptstyle M_J} \sum_{\scriptstyle M_J' \ne M_J}  (-1)^{2J-M_J-M'_J}  \\
&& \times \left(
\begin{array}{ccc} 
J & k &  J  \\
-M_J &  0 & M_J  
\end{array}
\right)
\left(
\begin{array}{ccc} 
J & k &  J  \\
-M'_J &  0 & M'_J  
\end{array}
\right) \nonumber \\ 
& & \times \sigma (\alpha J  M_J \to \alpha J  M'_J, v) \nonumber 
\end{eqnarray}
The expressions between parentheses denote $3j$-coefficients (Messiah \cite{Messiah}). After some calculations which are not detailed here, this equation can also be written as:
\begin{eqnarray} \label{eq9}
\sigma^k(\alpha  J, v) &=& \sigma (\alpha J, v)-(2k+1) \sum_{M_J,M'_J} 
(-1)^{2J-M_J-M'_J}  \nonumber \\
& & \times   \left(
\begin{array}{ccc} 
J & k &  J  \\
-M_J &  0 & M_J  
\end{array}
\right) 
\left(
\begin{array}{ccc} 
J & k &  J  \\
-M'_J &  0 & M'_J  
\end{array}
\right)  \\
&& \times  \sigma (\alpha J  M_J \to \alpha J  M'_J, v). \nonumber
\end{eqnarray} 
Where $\sigma(\alpha  J, v)$ is the elastic cross section for the level $\displaystyle | \alpha J \rangle$ and the relative velocity $v$:
\begin{eqnarray} \label{eq10}
\sigma (\alpha J, v)=  \sum_{M'_J}\sigma (\alpha J  M_J \to \alpha J  M'_J, v)
\end{eqnarray}
This cross section can also be written as an average over the initial states $M_J$ as usual:
\begin{eqnarray} \label{eq11}
\sigma (\alpha J, v)= \frac{1}{2J+1} \sum_{M_J,M'_J}\sigma (\alpha J  M_J \to \alpha J  M'_J, v)
\end{eqnarray}
where the angular average over all the directions of the 
collisions is  not necessary here because the averaging over $M'_J$ is equivalent to the angular average over the directions of the collision. This property will be used in Sect. \ref{sec6}, first line in equation (\ref{eq41}).

Since the depolarization cross section  $\displaystyle 
\sigma^k(\alpha  J, v)$ can be written as a linear combination of the $\displaystyle\sigma (\alpha J  M_J \to \alpha J  M'_J, v)$, the depolarization rate $\displaystyle 
D^k(\alpha  J, v)$  is  a linear combination of the collisional rates between the
Zeeman sublevels $\zeta(\alpha J M \to \alpha J M')$ (Sahal-Br\'echot \cite{Sahal2}). In particular we have (Sahal-Br\'echot \cite{Sahal2}), for $J=1$:
\begin{eqnarray} 
D^1(\alpha  1) & = & \zeta (10\to  1  \pm 1) + 2 \zeta
 (1  \mp 1 \to 1  \pm 1) \nonumber \\
D^2(\alpha  1)& = & 3 \zeta (1 0 \to1  \pm 1), \nonumber
\end{eqnarray}
and for $J=3/2$ (Gordeyev et al \cite{Gordeyev}; Roueff \& Suzor \cite{Roueff2}):
\begin{eqnarray}
D^1(\alpha\,\frac{3}{2}) & = & \frac{1}{5} \, \zeta (\frac{3}{2} \, \frac{-1}{2} \to \frac{3}{2} \, \frac{1}{2}) +\frac{2}{5} \, \zeta (\frac{3}{2} \, \frac{3}{2} \to \frac{3}{2} \, \frac{1}{2}) \nonumber \\
&+& \frac{8}{5} \, \zeta (\frac{3}{2} \, \frac{-3}{2}\to \frac{3}{2} \, \frac{1}{2})+\frac{9}{5} \, \zeta (\frac{3}{2} \, \frac{-3}{2} \to \frac{3}{2} \, \frac{3}{2}) \nonumber \\ 
D^2(\alpha\,\frac{3}{2}) & = & 2 \, \zeta (\frac{3}{2} \, \frac{3}{2} \to \frac{3}{2} \, \frac{1}{2}) +2 \, \zeta (\frac{3}{2} \, \frac{-1}{2} \to \frac{3}{2} \, \frac{1}{2}) \nonumber \\
 & + & 2 \, \zeta (\frac{3}{2} \, \frac{-3}{2} \to \frac{3}{2} \, \frac{1}{2}) \nonumber \\ 
D^3(\alpha\,\frac{3}{2}) & = & \frac{9}{5} \, \zeta (\frac{3}{2} \, \frac{-1}{2} \to \frac{3}{2} \, \frac{1}{2}) +\frac{8}{5} \, \zeta (\frac{3}{2} \, \frac{3}{2} \to \frac{3}{2} \, \frac{1}{2}) \nonumber \\
&+& \frac{2}{5} \, \zeta (\frac{3}{2} \, \frac{-3}{2} \to \frac{3}{2} \, \frac{1}{2})+\frac{1}{5} \, \zeta (\frac{3}{2} \, \frac{-3}{2} \to \frac{3}{2} \, \frac{3}{2}) \nonumber 
\end{eqnarray}
\section{Calculation of the cross sections} \label{sec3}
We will use a semi-classical treatment for the collision: we will assume that the hydrogen perturbing atom moves along a straight path, unperturbed by the 
collision process and characterized by an impact-parameter $b$.

 In the dyadic basis $|\alpha J M_J\rangle \langle  \alpha' J' M'_J|$ (or $|\alpha J F M_F\rangle \langle  \alpha' J' F' M'_F|$ if there is an hyperfine structure), the cross section follows from an integration over the impact-parameter $b$ and  averaging  the transition probability  $P(\alpha J M_J \to \alpha J M'_J, \vec{b}, \vec{v})$ over all the collision directions:
\begin{eqnarray}  \label{eq12} 
\sigma (\alpha J M_J \to \alpha J M'_J, v)& =&  \int_{0}^{\infty} 2\pi b db \Bigg(\frac{1}{8\pi^2}   \\
 &\times&  \oint P(\alpha J M_J \to \alpha J M'_J, \vec{b}, \vec{v})  d\Omega \Bigg) \nonumber   
\end{eqnarray}
and then the collisional transition rate is given by equation (\ref{eq3}).
The transition probability can be written in terms of the $S$-matrix
elements,
\begin{eqnarray} \label{eq13}   
P(\alpha J M_J \to \alpha J M'_J, \vec{b}, \vec{v}) && =  \\
&&|\langle \alpha J M_J| I - S(\vec{b}, \vec{v}) |\alpha J M'_J \rangle|^2 \nonumber
\end{eqnarray}  
where $I$ is the unit matrix and $T = I-S $ is the so-called transition
matrix. 
\subsection{Calculation of the S-matrix}
The Hamiltonian of the system atom-perturber can be written as: 
\begin{eqnarray} \label{eq14}
H=H_A + H_P + V,
\end{eqnarray}
where $H_A$ is the atomic Hamiltonian, $H_P$ that of the perturber and $V$
is the interaction energy. $H_A$ can be written as: 
\begin{eqnarray} \label{eq15}
H_A=H_0 + H_{FS} + H_{HFS},
\end{eqnarray}
where $H_0$ contains the electrostatic part, $H_{FS}$ the fine structure and $H_{HFS}$ the hyperfine structure if it exists.

It can be shown that the hyperfine splitting $\Delta E_{HFS}$ is always small when compared to $V$ and
thus $J$ and $I$ (the nuclear spin) can be decoupled during the collision process. This arises from the Heisenberg uncertainty principle: if  $\tau$ is a typical  collision time, one can ignore the hyperfine structure if 
\begin{eqnarray} \label{eq16}
\tau \; \Delta E_{HFS} \ll 1,
\end{eqnarray}
where $\displaystyle \tau \sim \frac{b}{v}$,  $b$ being a typical impact-parameter (of the order of $10 a_0$, $a_0$ being the Bohr radius) and $v \sim 
4.8  \times 10^{-3} \textrm{a.u.} $ for a temperature of about 5000 K so that $\tau \sim 2.08  \times  10^3 \textrm{a.u.}$. In the case of the Na atom the nuclear spin  is $I=3/2$, which yields two 
hyperfine levels in the ground state $3s \;^2S_{1/2}$, $F=1$ and $F=2$. For 
the excited $3p\; ^2P_{3/2}$ state there are four hyperfine levels: $F=0$, $F=1$, 
$F=2$ and $F=3$.  The energy differences between these levels  are given in 
Table \ref{NaHFS}. 
In the case of the ground state $3s\; ^2S_{1/2}$, $\displaystyle \Delta 
E_{HFS} \simeq  2.694 \times 10^{-7} \textrm{a.u.}$ $(1772 \;\textrm{MHz})$ 
(Ackermann \cite{Ackermann}) and then  $\displaystyle \tau \; \Delta 
E_{HFS} \simeq 5.6 \times 10^{-4} \ll 1$. For the excited level the most 
important  hyperfine splitting is, of course, between the levels $F=0$ and 
$F=3$ where $\displaystyle \Delta E_{HFS} \simeq  1.653 \times 10^{-8} 
\textrm{a.u.}$ $(108.72 \;\textrm{MHz})$  (Gangrsky et al \cite{Gangrsky}) 
and then  $\displaystyle \tau \; \Delta E_{HFS} \simeq 3.4 \times 10^{-5} \ll 1$.

The hyperfine splitting can always be ignored in the treatment of the collision  under these conditions. We have $\vec{F}=\vec{I}+\vec{J}$ and $M_F=M_I+M_J$, and
\begin{eqnarray} \label{eq17}
\displaystyle |\alpha J F M_F \rangle = \sum_{M_I,M_J} \langle I J M_I M_J| F M_F\rangle |J M_J \rangle | I M_I \rangle,
\end{eqnarray}
where $\langle I J M_I M_I| F M_F\rangle$ is a Clebsch-Gordan coefficient. Then the matrix element 
\begin{eqnarray} \label{eq18}
\langle \alpha J F M_F |S|\alpha' J' F' M_F' \rangle =  \sum_{M_I,M_J,M_I',M_J'} \langle I J M_I M_J| F M_F\rangle \nonumber \\
\times  \langle I J' M_I' M_J'| F' M_F'\rangle  
 \langle \alpha J  M_J|\langle I  M_I| S | I  M_I' \rangle |\alpha' J'  M_J'\rangle.\quad
\end{eqnarray} 
If the hyperfine structure can be neglected during the collision, the $S$-matrix is diagonal in $I$ and its elements do not depend on $M_I$ ($I$ is conserved). Thus $M'_I = M_I$ and 
\begin{eqnarray} \label{eq19}
\langle \alpha J F M_F |S|\alpha' J' F' M_F' \rangle  =  \sum_{M_I,M_J,M'_J} \langle I J M_I M_J| F M_F\rangle \nonumber \\
\times \langle I J' M_I M_J'| F' M'_F\rangle  
\langle \alpha J  M_J| S | \alpha' J'  M_J'\rangle \quad
\end{eqnarray}
This relation holds for each impact-parameter and each relative velocity.

We recall the relation between Clebsch-Gordan coefficients and  
$3j$-coefficients:\\
$\langle j_1 j_2 m_1 m_2| j m\rangle = \sqrt{2j+1} (-1)^{j_1-j_2+m}
\left(
\begin{array}{ccc} 
j_1 & j_2 &  j  \\
m_1 &  m_2 & -m  
\end{array}
\right)
$\\
We use the convention of  Messiah (\cite{Messiah}) for the phase factors.
\subsection{Calculation of the $S$-matrix elements between fine sublevels}
The fine structure splitting can generally be neglected during the collision 
problem at the temperature of interest in the second solar spectrum ($\sim 5000
 K$). The validity of the condition is the same as above, except that the fine
 structure splitting is larger than the hyperfine splitting.
For example in the case of the Na atom, at $5000 K$, $\Delta E_{FS}\simeq 0.78  \times 10^{-4} \textrm{a.u.}$ and $\tau \simeq 2.08 \times 10^3 \textrm{a.u}$ and so $\tau \;  \Delta E_{FS} \simeq 0.16 < 1$.
This has been discussed in earlier papers focussed on collisional line 
broadening, see for example Roueff (\cite{Roueff2}) and related papers. 
She showed that the fine structure splitting is of importance at low temperatures (200-500 K) 
for the sodium $D$ lines and negligible at higher temperatures for line 
broadening studies (500 K and higher). The $D_1$ and $D_2$ lines have in fact
the same collisional width for hydrogen perturbers at solar temperatures.

If the fine structure is neglected, where $\vec{J}=\vec{S}+\vec{l}$, and $M_J=M_S+m_l$, we have:   
\begin{eqnarray} \label{eq20}
\langle \alpha l J M_J |S|\alpha' l' J' M'_J  \rangle =\displaystyle \sum_{M_S,m_l,m'_l} 
\langle S l M_S m_l| J M_J\rangle \\
\times \langle S l' M_S m'_l| J' M'_J\rangle  
\langle \alpha l  m_l| S  |\alpha' l'  m'_l\rangle, \nonumber 
\end{eqnarray}
\textrm{cf. equation (\ref{eq19}}). 

However, we notice that the parallelism between line broadening and depolarization cross sections must be used carefully in the peculiar following case. Depolarization of hyperfine components of spherically 
symmetric $J=J'= 1/2$ levels (ground state of sodium for instance), needs the matrix elements $M_J=\pm \frac{1}{2} \to M'_J=\pm \frac{1}{2}$. If the fine structure is neglected, they are  zero and the depolarization 
 and polarization transfer rates of the hyperfine components of the ground states are zero. Kerkeni et al (\cite{Kerkeni1}) calculated the hyperfine collisional depolarization rates
for the ground state of sodium with a quantum chemistry method which takes 
into account the fine structure splitting during the interaction processes. They showed that these rates are of the same order of magnitude as those 
of upper levels for which the fine structure splitting plays a minor role in
the collisional processes. In fact this remark does not concern $l \ge 1$ states and is very specific. However, we consider that it is important to mention it.
\section{The model of Anstee, Barklem and O'Mara (ABO)} \label{sec4}
The present semi-classical model used for the depolarization  is the same one 
as that already developed for  line broadening theory applications in the 
1990's by Anstee, Barklem and O'Mara (Anstee \cite{Anstee2}; Anstee \& 
O'Mara \cite{Anstee1}, \cite{Anstee3}; Anstee et al. \cite{Anstee4}; Barklem \cite{Barklem3};  Barklem \& O'Mara \cite{Barklem1}, Barklem et al. \cite{Barklem2}).
\begin{figure}[htbp] 
\begin{center}
\includegraphics[width=8 cm]{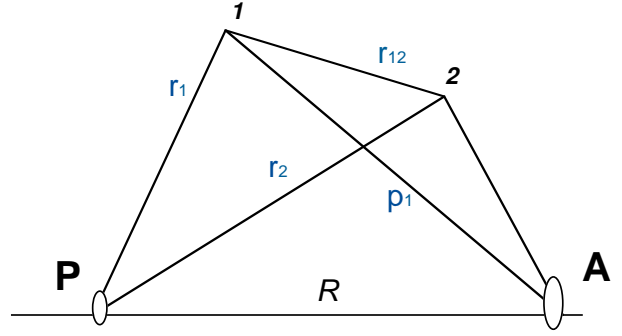}
\end{center}
\caption{The perturbed atom core is located at A and the hydrogen perturbing 
core (a proton) at P. Their valence electrons are denoted by 1 and 2 
respectively.}
\label{perturbed}
\end{figure}

The theory is  described in these papers, and thus only a brief review
is presented in the following:
  
1. The model is of semi-classical nature, the  internal states of the two
 atoms are 
treated within the framework of quantum mechanics and the collision is  
modelled with the straight path classical approximation: it is assumed that 
the interparticle interaction is too weak to deflect the perturbed particle 
from the straight classical path. 
   
2. The interaction energy between the two atoms is smaller than the energy 
eigenvalues of the isolated atoms, so that  the Rayleigh-Schr\"odinger perturbation theory can be used for obtaining perturbed eigenvalues. Ionic
and exchange interactions are neglected. The hyperfine and fine structure 
interactions are also neglected.

3. Time-independent perturbation theory is applied  to second order. It is assumed that the 
energy denominator in the second order term
can be replaced by a suitable average energy $E_p$. This is  
the Uns\"old approximation (Uns\"old \cite{Unsold1}; Uns\"old \cite{Unsold2}). The value $E_p=$ -4/9, in atomic units, is adopted. This approximation works  
when the  energy level separations of the perturbed atom are small compared 
with those of hydrogen. For neutral atoms this condition is 
generally satisfied. This is particularly valid when considering interactions 
with hydrogen in the ground state, like in this work, which is well separated 
from any other levels (Anstee \& O'Mara \cite{Anstee1}).

4. Unperturbed atomic eigenfunctions are used. The Coulomb approximation with
quantum defect is used.
\section{Treatment of the collision} \label{sec5}
Each collision is characterized by an impact-parameter vector $\vec{b}$ and a 
relative velocity vector $\vec{v}$. Since the hyperfine and fine structure 
interactions are neglected, equation (\ref{eq14}) becomes: 
\begin{eqnarray} \label{eq21}
H=H_0+H_P+V 
\end{eqnarray}
where $V$ is the interaction energy which is assumed to be totally 
electrostatic.
Only $V$ depends on time.
In the coordinate system of Fig. \ref{perturbed}, $V$ is  given, in atomic units, by (Anstee \& O'Mara \cite{Anstee1}):
\begin{figure}[htbp]
\begin{center}
\includegraphics[width=8 cm]{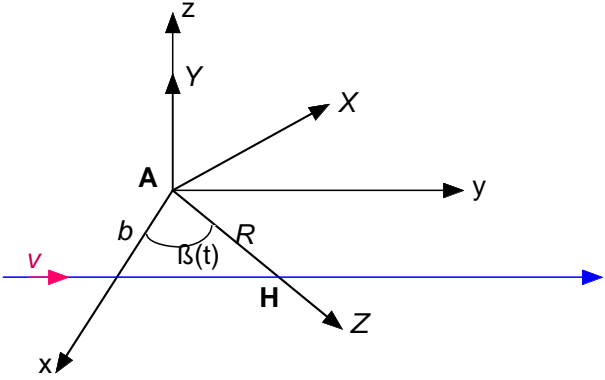}
\end{center}
\caption{Trajectory of the perturber (H) in the two reference frames. $R$ is the interatomic separation, $R=\sqrt{b^2+v^2t^2}$. $Axyz$ and $AXYZ$ are the atomic  and   rotating frames.}
\label{trajectory}
\end{figure}
\begin{eqnarray} \label{eq22}
V=\frac{1}{R}+\frac{1}{r_{12}}-\frac{1}{r_2}-\frac{1}{p_1}
\end{eqnarray}
and atomic units are used hereafter.

Since the hydrogen perturber remains in its ground state $1s$ during the 
collision, the wave function $|\psi \rangle$ of the system (atom+perturber), 
 is the product of the wave functions $|\psi \rangle_{\scriptstyle (A)}$ of
the perturbed atom and  that of hydrogen in its ground state $|1 0 0 \rangle_{\scriptstyle (H)}$:
\begin{eqnarray} \label{eq23}
|\psi \rangle =|\psi \rangle_{\scriptstyle (A)}|1 0 0 \rangle_{\scriptstyle (H)}.
\end{eqnarray} 
The eigenenergy of the hydrogen perturber in its ground state is denoted by 
$ E^0_{H}$. Following Anstee \& O'Mara (\cite{Anstee1}), we use  second-order 
time-independent perturbation theory for the eigenenergy of 
the system. 

We also use the unperturbed atomic wave functions defined by:
\begin{eqnarray} \label{eq24}
H_0 |n l m_l\rangle = E^0_{n l m_l}|n l m_l\rangle; |\psi \rangle_{\scriptstyle (A)} = |n l m_l\rangle.
\end{eqnarray}

They form a complete set of eigenfunctions $|n l m_l\rangle$ and eigenvalues  
$E^0_{n l m_l}$.
The perturbed eigenvalues of the system of Hamiltonian $H$ can be written as: 
\begin{eqnarray} \label{eq25}
E_{i}  =  E^0_i + \langle i | V |i\rangle + \sum_{j\ne i} \frac{ \langle i | V 
| j \rangle  \langle j |V | i \rangle }{E^0_i - E^0_j},
\end{eqnarray}
the summation being over all product states $|j\rangle= | \psi 
\rangle_{(A)}|\psi  \rangle_{(H)}$ except $|i\rangle= | n l m_l 
\rangle_{(A)}|100  \rangle_{(H)}$, and where $\displaystyle E^0_i = E^0_{n l m_l}+E^0_{H}$.

Quenching is neglected and thus we consider only the subspace $nl$ (2$l$+1 
states) and we denote the 
product state of the two separated atom at $R=\infty$  by $|M_l\rangle$ (Anstee \& O'Mara \cite{Anstee1}),
\begin{eqnarray} \label{eq26}
|M_l \rangle = |i\rangle=|n l m_l \rangle_{\scriptstyle (A)}|1 0 0 
\rangle_{\scriptstyle(H)}.
\end{eqnarray} 
We introduce an effective potential $V_{eff}$, the diagonal elements 
of
which are defined as 
\begin{eqnarray} \label{eq27}
\langle M_l | V_{eff} | M_l \rangle =E_{M_l} - E^0_{M_l} 
\end{eqnarray}

The evaluation of the second order contribution to the above expression (\ref{eq25}) is very difficult because it requires a summation over the
complete set of eigenstates of the two isolated atoms. However, the  expression  can be greatly simplified by using the Uns\"old approximation, i.e. by replacing $ \displaystyle E^0_{i} - E^0_{j}$ by the constant value $\displaystyle E_p$. With this approximation, the expression for $\displaystyle V_{eff}$ becomes 
\begin{eqnarray} \label{eq28}
\langle M_l \big\vert V_{eff}\big\vert M_l \rangle & = & \langle 
M_l \big\vert V\big\vert M_l \rangle - \frac{1}{E_p}(\langle M_l \big\vert V\big\vert M_l \rangle )^2  \nonumber \\
&&+  \frac{1}{E_p} \langle M_l \big\vert V^2\big\vert M_l \rangle, 
\end{eqnarray} 
where $\displaystyle \langle M_l | V_{eff} | M_l \rangle$ is the 
so-called  Rayleigh-Schr\"odinger-Uns\"old (RSU) potential which is used throughout
the ABO model. Then, for computing the RSU potential, explicit expressions for the wave functions are needed.

 The perturbed atom is modelled as a single valence electron outside a spherical ionic core. The chosen radial functions are of Coulomb type (Seaton \cite{Seaton}). The quantum defect and thus the effective quantum number
$ n^*=[2 (E_\infty - E_{n l })]^{-1/2}$ is a fundamental ingredient of the calculation of the interaction potential used in the ABO model. We refer to the ABO papers for more details concerning the interaction potential.

The $S$-matrix  requires the time-dependent Schr\"odinger equation to be solved, i.e.
\begin{eqnarray} \label{eq29}
(H_0+H_P+V_{eff}) \big | \psi(t) \rangle =\textrm{i} \frac{d\big | \psi(t)\rangle}{dt}
\end{eqnarray}
where the interaction potential $V$ is now replaced by $V_{eff}$. The $S$-scattering matrix is defined by:
\begin{eqnarray} \label{eq30}
S = U(+\infty, -\infty)
\end{eqnarray}
where $U(+\infty, -\infty)$ is the evolution operator in the interaction representation, where
\begin{eqnarray} \label{eq31}
\widetilde{\psi}(t+s)=U^{\dag} (t+s, t) \psi(t) = \textrm{e}^{\textrm{i} (H_0+H_P) s} \psi(t), 
\end{eqnarray}
is the corresponding wave function. The wave function is expressed in terms of the basis formed by the eigenvectors $ | M_l \rangle$ of $(H_0+H_P)$ given by (\ref{eq26}) so that:
\begin{eqnarray}  \label{eq32}
\big | \psi (t) \rangle = \displaystyle \sum_{M_l} a_{M_l}(t)   \textrm{e}^{-\textrm{i} {E^0_{M_l} t}} \big | M_l \rangle;&&  \\
&& \big | \widetilde{\psi} (t) \rangle = \displaystyle \sum_{M_l} a_{M_l}(t)    \big | M_l \rangle  \nonumber
\end{eqnarray}
and we obtain a system of $(2l+1)$ coupled linear differential equations of first order: 
\begin{eqnarray} \label{eq33}
\textrm{i}  \frac{\partial a_{M_l}(t)}{\partial t}=\sum_{M_k=-l}^l  a_{M_k}(t) 
\langle   M_l \big\vert V_{eff}\big\vert  M_k \rangle.   
\end{eqnarray}
We have to solve this system for the initial conditions:
\begin{eqnarray} \label{eq34}
a_{M_l}(-\infty) = \delta(M_i,M_l),
\end{eqnarray}
where $\displaystyle\delta$ is the Kronecker symbol and
\begin{eqnarray} \label{eq35}
\langle   M_l| S | M_i \rangle= \langle   M_i| S | M_l \rangle=a_{M_l}(+\infty).
\end{eqnarray}
Thus the system has to be solved $(2l+1)$ times for obtaining all the elements
of the $S$-matrix for the $(n l)$ subspace.

For the calculations, two frames of reference are used. The first one is the  atomic frame in which the perturbed atom is 
stationary at the origin, the  quantization axis is taken as 
perpendicular to the 
collision plane  $( $\vec{b}$, $\vec{v}$)$ and the second one is the rotating 
or molecular frame in which the interatomic axis is taken as the quantization 
axis. These reference frames are shown in Fig. \ref{trajectory}.
This choice of frames gives the simplest relationship between states quantized
relative to the atomic and the rotating frames respectively (Roueff \cite{Roueff1}). 

 The   interaction  potential $V$  has a cylindrical symmetry about the 
internuclear axis $(OZ)$ and so $V_{eff}$ is  diagonal in the  rotating frame.

 A transformation of the angular momentum states from the atomic basis to
the rotating frame is achieved by using the rotations matrices  
(Messiah \cite{Messiah}). The Euler angles are given by $(\beta, \frac{\pi}{2}, \frac{\pi}{2})$ and equation (\ref{eq33}) becomes       
\begin{eqnarray} \label{eq36}
\textrm{i}  \frac{\partial a_{M_l}(t)}{\partial t}=\sum_{M_k M'_k} 
\mathcal{D}^{(l)}_{M_lM'_k} \mathcal{D}^{\dag (l)}_{M_lM'_k}a_{M_k}(t) \langle M_l \big\vert V_{eff}\big\vert M_l \rangle
\end{eqnarray}
 when $\cal{D}$ is the rotation operator.

 For a given quantum number $l$ we obtain $(2l+1)$ coupled differential
equations. These coupled equations must be solved for obtaining
the $ (2l+1)^{2} \; S$-matrix elements.
 
 The interaction matrix is Hermitian and the $S$-matrix is unitary and symmetric. If
the quantization axis is perpendicular to the collision plane the selection 
rules are: $ \Delta {m_l}= \pm2, \pm4 ...$. This rule will disappear when we 
 take an average over all possible orientations of the collision plane, since
 the collisions are isotropic.
\section{Depolarization calculations for $l=1$} \label{sec6}
Let us consider collisions between an atom in  an $l=1$  state
and  hydrogen in its ground state $1s$. The state of the system during the collision becomes explicitly: 
\begin{eqnarray} \label{eq37}
\big | \psi (t) \rangle & = & a_{1}(t) \big | 11 \rangle \big | 00 
\rangle \textrm{e}^{-\textrm{i}\; E_{1}^0\; t} + a_{0}(t) \big | 10 \rangle 
\big | 00 \rangle \textrm{e}^{-\textrm{i}\; E_0^0 \;t } \nonumber \\
&+&a_{-1}(t) 
\big | 1-1 \rangle 
\big | 00 \rangle \textrm{e}^{-\textrm{i}\; E_{-1}^0 t}
\end{eqnarray} 
and the coupled differential equations become (Anstee \& O'Mara 
 \cite{Anstee1}):
\begin{eqnarray} \label{eq38}
\textrm{i} \frac{\partial a_{1}(t)}{\partial t} & = &a_{1}(t) V_+ + \textrm{e}^{2\textrm{i}\beta} a_{-1}(t) V_- \nonumber \\
\textrm{i} \frac{\partial a_{0}(t)}{\partial t} & = & a_{0}(t) \langle 1 |V_{eff} | 1\rangle  \\
\textrm{i} \frac{\partial a_{-1}(t)}{\partial t} & = & a_{-1}(t) V_+ + \textrm{e}^{-2\textrm{i}\beta} a_{1}(t) V_- \nonumber  
\end{eqnarray}
where:
\begin{eqnarray} \label{eq39}
V_{\pm}=\frac{1}{2}(\langle 1 |V_{eff} | 1\rangle \pm \langle 0 |V_{eff} | 0
\rangle).
\end{eqnarray}
The equations are integrated with a Runge-Kutta-Merson routine and then we 
obtain the transition matrix elements  in the $| \alpha l m_l \rangle$ basis 
for a given velocity and  impact-parameter.

We recall that we need, for depolarization calculations, the $T$-matrix elements in the $| \alpha J M_J \rangle$ basis. Since the spin is  neglected, the $S$-matrix is diagonal in $S$ and its elements do not depend on $M_S$. Following equation (\ref{eq20})  we   obtain:    
\begin{eqnarray} \label{eq40}
\langle \alpha J M_J|T| \alpha J M'_J \rangle & =&\sum_{m_l,m'_l,M_{S}} (-1)^{2S-2l+M_J+M'_J}  (2J+1) \nonumber \\
&\times &\left(
\begin{array}{ccc} 
S & l &  J  \\
M_S &  m_l & -M_J  
\end{array}
\right) \\
&\times & \left(
\begin{array}{ccc} 
S & l &  J \\ 
M_S &  m'_l & -M'_J  
\end{array}
\right)
 \langle \alpha  l  m_l | T |\alpha  l  m'_l \rangle \nonumber
\end{eqnarray}
The matrix elements $\langle \alpha J M_J|T| \alpha J M'_J \rangle $ are obtained in a basis where the quantization 
axis is perpendicular to the collision plane. We have now to perform an average 
 of the transitions probabilities over all possible orientations since collisions are isotropic. This angular
average is denoted by $\langle \quad \rangle_{av}$.

Using formula (16) by Sahal-Br\'echot (\cite{Sahal1}), it can be shown that, in the irreducible tensorial operator basis, the angular average of the depolarization transition probability  is given by:
\begin{eqnarray}  \label{eq41} 
\langle P^k(\alpha J, b, v) \rangle_{av}  = \frac{1}
{2J+1} \sum_{\mu,\mu'}|\langle \alpha \; J \; \mu |T|\alpha \; J \; \mu'\rangle|^2 \nonumber \\
-\sum_{\mu,\mu' , \nu,\nu'} \langle \alpha \; J \; \mu |T|\alpha \; J \; \mu'\rangle  \langle \alpha \; J \; \nu |T|\alpha \; J \; \nu'\rangle^* \quad \quad \quad \\
\times \; \sum_{\chi} (-1)^{2J+k+\mu-\mu'}
\left(
\begin{array}{ccc} 
J & J &  k  \\
-\nu' &  \mu' & \chi  
\end{array}
\right)
\left(
\begin{array}{ccc} 
J & J &  k \\ 
\nu &  -\mu & -\chi  
\end{array}
\right) \nonumber
\end{eqnarray}
Owing to the 
selection rules on the $3j$-coefficients, the summation over $\chi$ is reduced to a single term, since $\chi = -(\mu'-\nu')=-(\mu-\nu)$.

The depolarization cross section may then  be computed from: 
\begin{eqnarray} \label{eq42}
\sigma^k(\alpha J, v)=2 \pi \int_{0}^{\infty} \langle P^k(\alpha J, b, v) \rangle_{av} \; b \; db
\end{eqnarray}
At small impact-parameters, the $\langle P^k(\alpha J, b, v) \rangle_{av}$ values become highly oscillatory, and it is impractical
to integrate numerically. In addition, at these separations the perturbation 
theory  breaks down and the calculated RSU potentials become invalid. 
Thus, an impact-parameter cutoff is used (Anstee \& O'Mara \cite{Anstee1}): 
\begin{eqnarray} \label{eq43}
\sigma^k(\alpha J, v) \simeq \pi b_0^2 +2 \pi \int_{b_0}^{\infty} \quad
\langle P^k(\alpha J, b, v) \rangle_{av} \; b \; db
\end{eqnarray} 
Integration over the velocity distribution for a temperature $T$ can be performed to obtain the depolarization rate which is given by: 
\begin{eqnarray} \label{eq44}
D^k(\alpha  J, T) &\simeq& n_H \int_{0}^{\infty}v f(v) dv  
 \Bigg( \pi b_0^2  \\
&&+ 2 \pi \int_{b_0}^{\infty}  \langle  P^k(\alpha J, b, v)  \rangle_{av} 
 \;b\; db \Bigg) \nonumber
\end{eqnarray}
where $b_0$ is the cutoff impact-parameter and here  $b_0=3 a_0$ as 
in Anstee \& O'Mara (\cite{Anstee1}) (see Sect. \ref{sec8}).

\section{Results}  \label{sec7}
\subsection{Our results for $^1P_1$ and $^2P_{3/2}$ levels}
The perturbed atom is modelled as a single valence electron outside a 
spherical ionic core. If the energy of the state $|\alpha l  \rangle$ of the 
valence electron is denoted by $E_{\alpha l }$ and the binding energy of the 
ground state by $E_\infty$,  the binding energy of the level $|\alpha l  
\rangle$ is ($ E_\infty - E_{\alpha l}$) and is related to the effective principal 
quantum number by $n^*=[2 (E_\infty - E_{n l })]^{-1/2}$.

For example: consider the case of the neutral calcium line $4227 \AA$ 
(transition $4s^2-4s4p$). The energy of the $4 p$ level   and the binding 
energy are 
$\simeq 0.1077 \textrm{a.u.} (\sim 23652 \; \textrm{cm}^{-1})$  and  $\simeq 0.2245 \textrm{a.u.}  ( 49305.72 \; \textrm{cm}^{-1})$ respectively (Wiese et al \cite{Wiese}). Hence the effective principal quantum number is $n^* \simeq 2.077$.

Note that if the parent configuration of the level is not the same as that of the ground state, this should be accounted for by using the appropriate series limit for the excited parent configuration instead of the ground state binding energy. A particular example was explained in detail by Barklem, Anstee \& O'Mara (\cite{Barklem3p}) (see also Barklem, Piskunov \& O'Mara \cite{Barklem4}). As in Anstee \& O'Mara (\cite{Anstee1}), we can calculate the cross sections as a function of $n^*$.
Table \ref{cross} gives   $\sigma^k(\alpha 1, v)$ and $\sigma^k(\alpha \frac{3}{2}, v)$ as a function of $n^*$ for a typical relative velocity  $v=10 \; \textrm{km} \; \textrm{s}^{-1}$.

\begin{table}
\begin{center}
\begin{tabular}{|l|c|c|c|c|r|}
\hline
$n^*$ & $\sigma^1(\alpha 1)$ & $\sigma^2(\alpha 1)$ & $\sigma^1(\alpha \frac{3}{2})$ & $\sigma^2(\alpha \frac{3}{2})$ & $\sigma^3(\alpha \frac{3}{2})$ \\
\hline
$1.5$ &203&192&	79&	141&	134\\
\hline 
$1.6$ &233&228&90&	165&	157\\
\hline 
$1.7$ &	276&258&	100&	187&	176\\
\hline 
$1.8$&	323&301&	114&	217&	203\\
\hline
$1.9$&	377&349&	130&	252&	233\\
\hline 
2&438	& 405	&147	&290	&268\\
\hline
$2.1$&507	 &466	&167	&334	&307\\
\hline 
$2.2$ &581	&533	&189	&381&	349\\
\hline 
$2.3$ &663	&605	&212	&432	&393\\
\hline 
$2.4$ &751	&683	&237	&487	&442\\
\hline 
$2.5$ &849	&769	&266	&549	&496\\
\hline 
$2.6$ &949	&869	&295	&616&	562\\
\hline 
$2.7$ &	1058&970	&327	&685	&627\\
\hline 
$2.8$ &1180	&1075	&362	&761&691\\
\hline 
$2.9$ &1299	&1176	&396	&835	&753\\
\hline 
$3$ &1423	&   1282	&431	&911	&817\\ 
\hline  
\end{tabular}
\end{center}
\caption{Variation of the depolarization cross section, for a 
relative velocity  of $10 \; \textrm{km} \; \textrm{s}^{-1}$,  
with  effective principal quantum number. Cross sections are in 
atomic units.}
\label{cross}
\end{table}
We thus can obtain the depolarization 
cross section for any level of any atom  by interpolation of the 
values given in Table \ref{cross} for the appropriate $n^*$ value for the state of interest. We notice that the dependence of cross sections  on velocity is very close to
\begin{eqnarray} \label{eq45}
\sigma^k(\alpha J, v)= \sigma^k(\alpha J, v_0)(\frac{v}{v_0})^{-\lambda^k(\alpha J)} ,
\end{eqnarray}
where ${v_0}$ is the velocity at which the cross section is known ($10 \; \textrm{km} \; \textrm{s}^{-1}$). Table \ref{lambda^k} shows the so-called velocity exponent $\lambda^k(\alpha J)$ (Anstee \& O'Mara \cite{Anstee1}).  In order to obtain the depolarization cross section for  any velocity $v$ from Tables \ref{cross} and \ref{lambda^k}  equation (\ref{eq45})  can be used.

\begin{figure}[htbp]
\begin{center}
\includegraphics[width=8 cm]{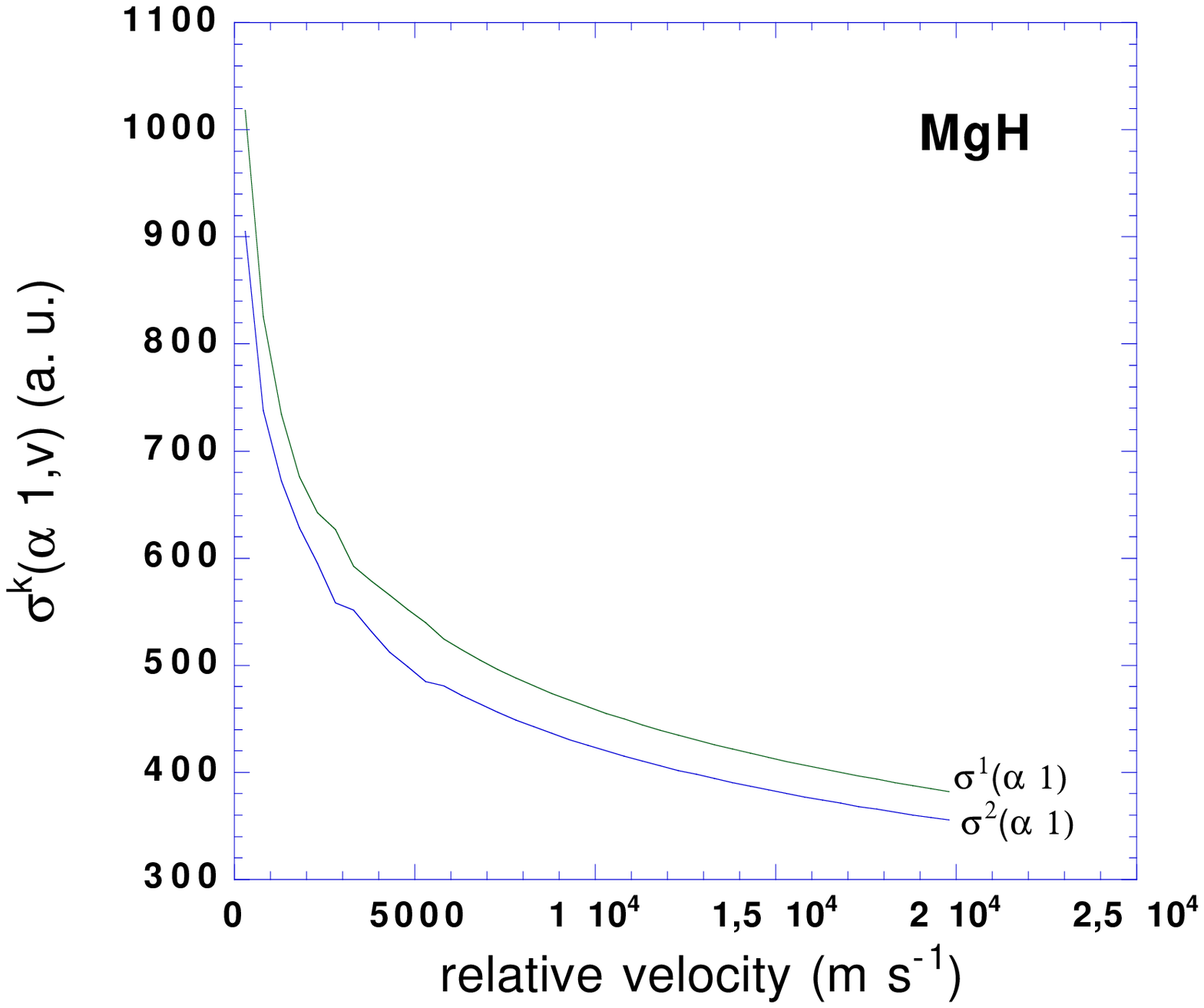}
\includegraphics[width=8 cm]{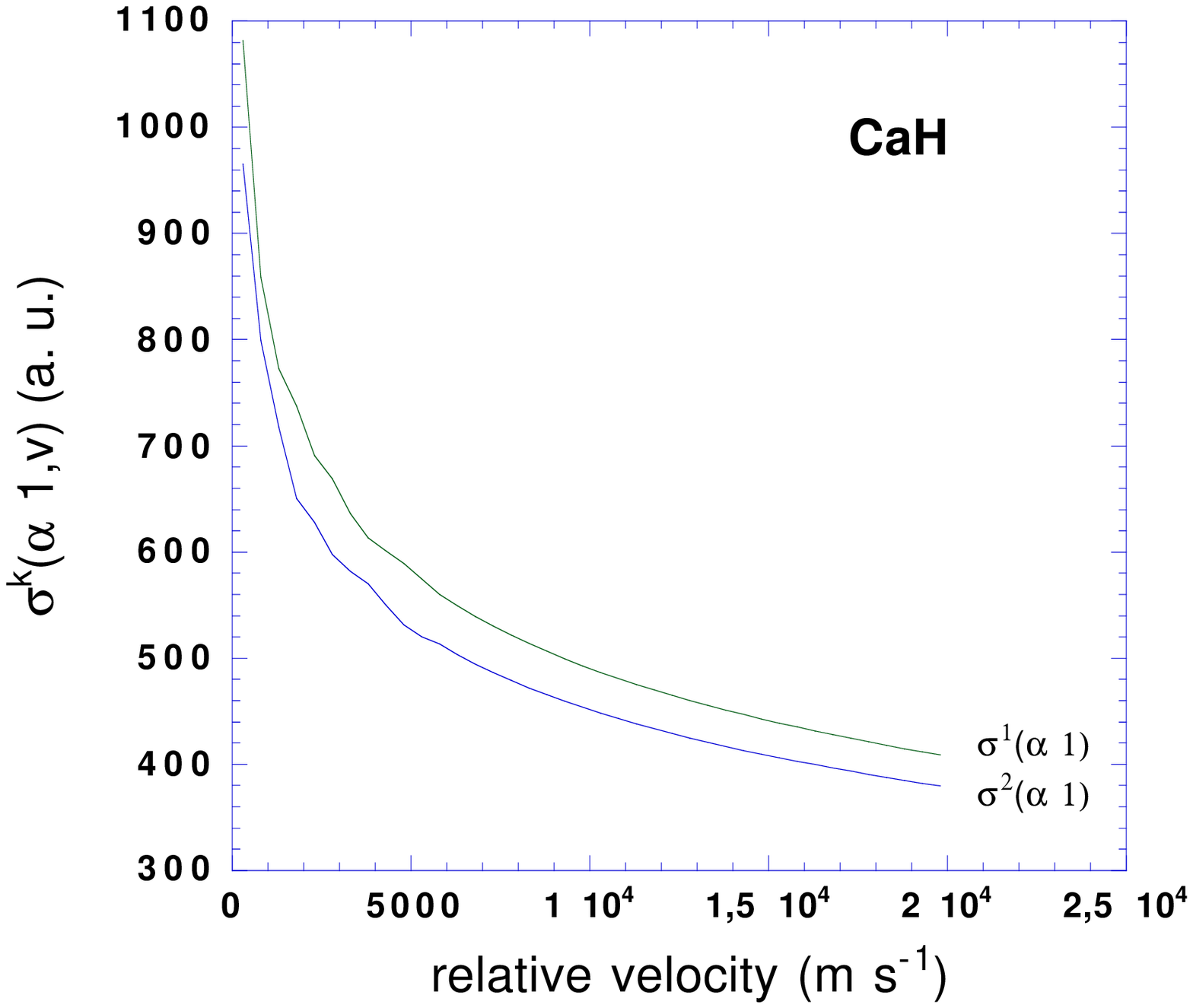}
\includegraphics[width=8 cm]{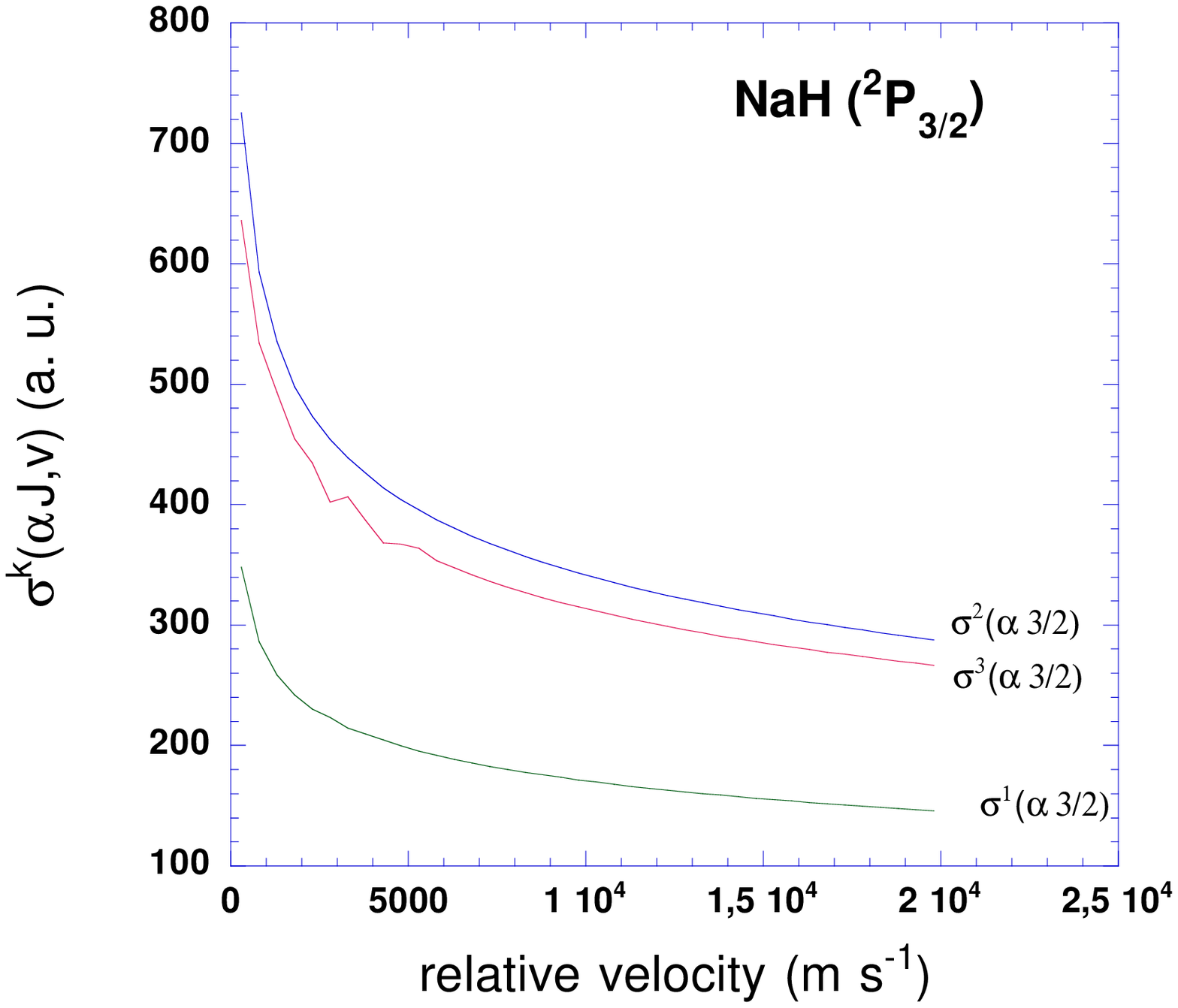}
\end{center}
\caption{Depolarization cross sections as a function of relative velocity.}
\label{depolarization}
\end{figure}

\begin{figure}[htbp]
\begin{center}
\includegraphics[width=8 cm]{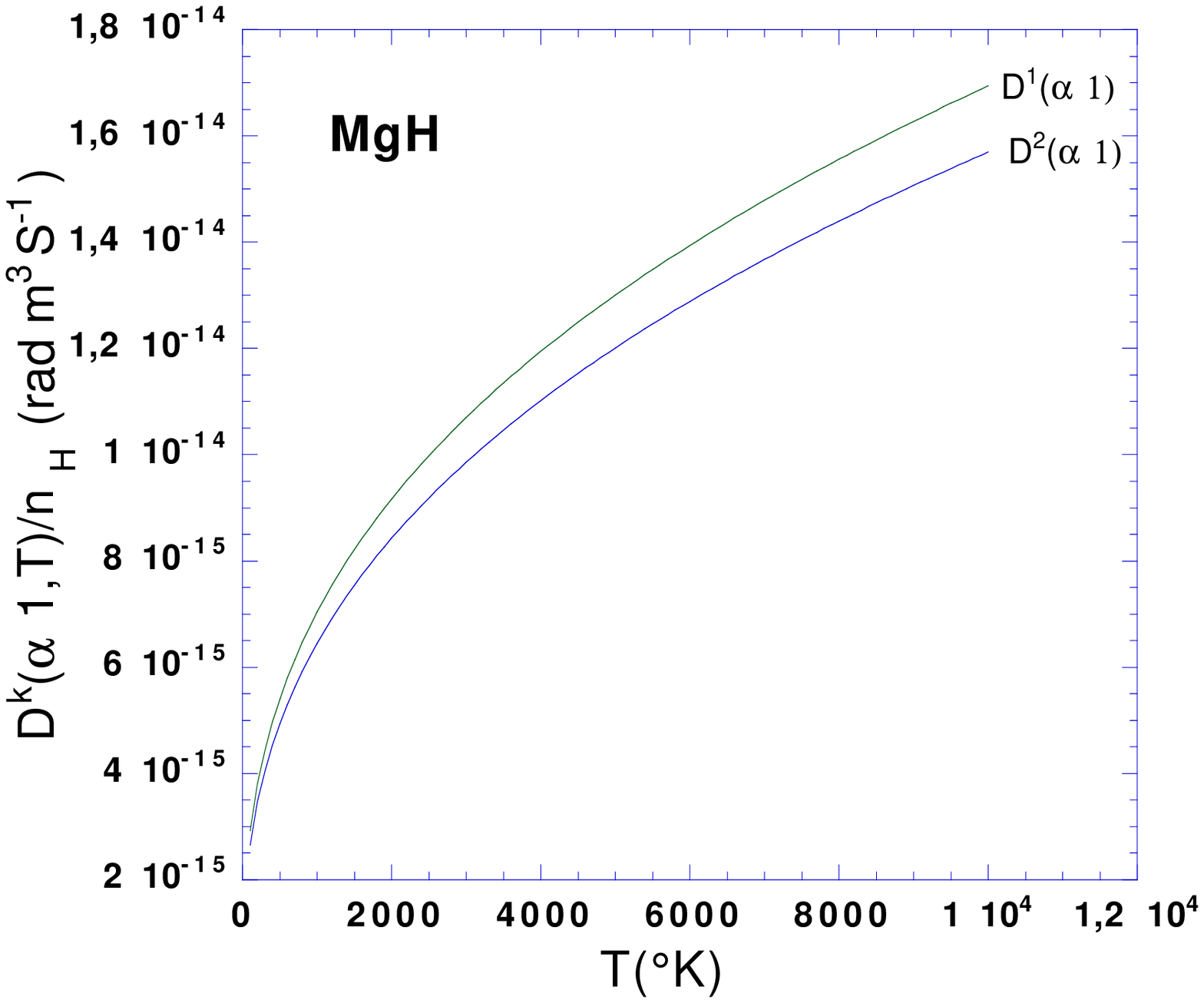}
\includegraphics[width=8 cm]{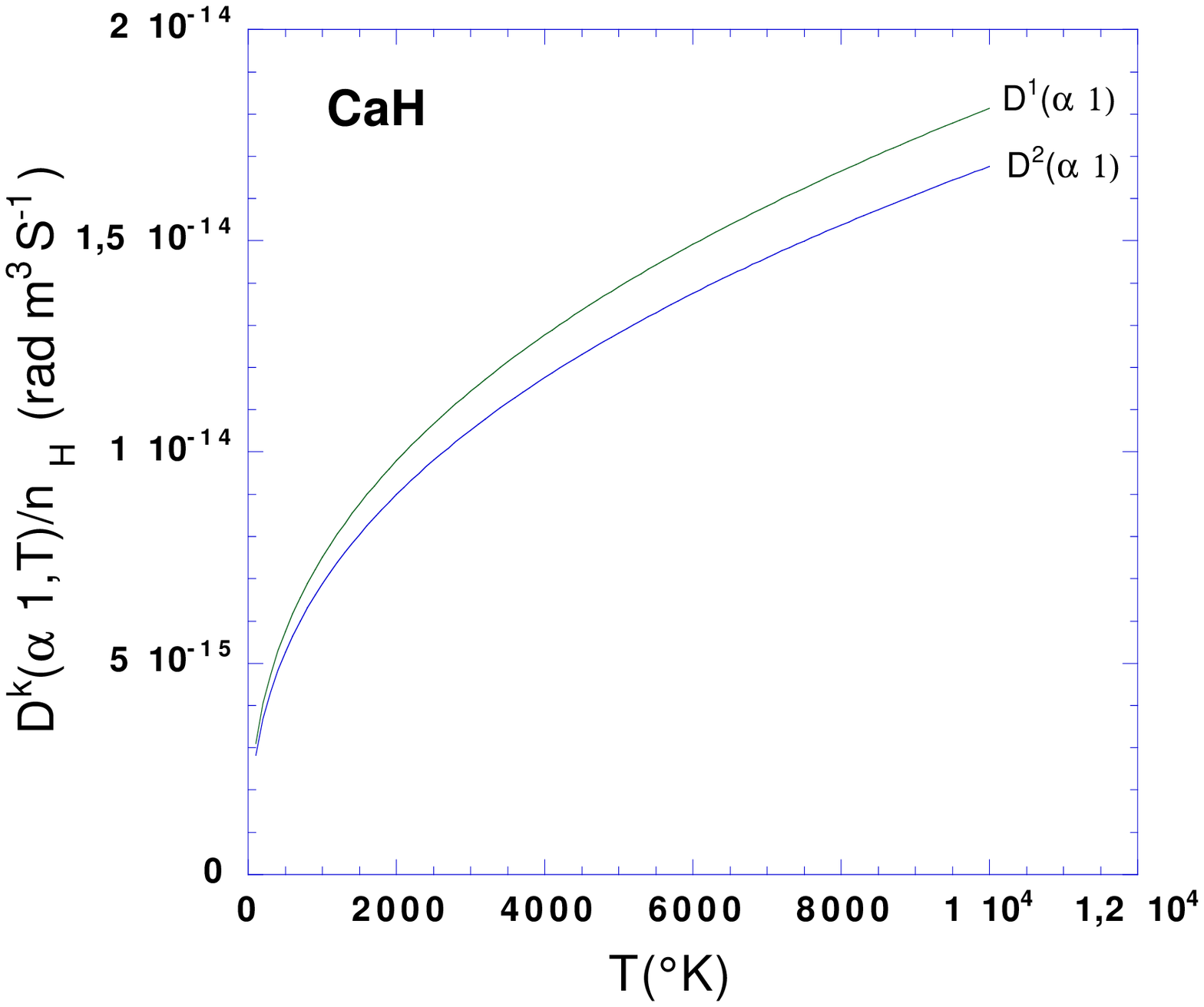}
\includegraphics[width=8 cm]{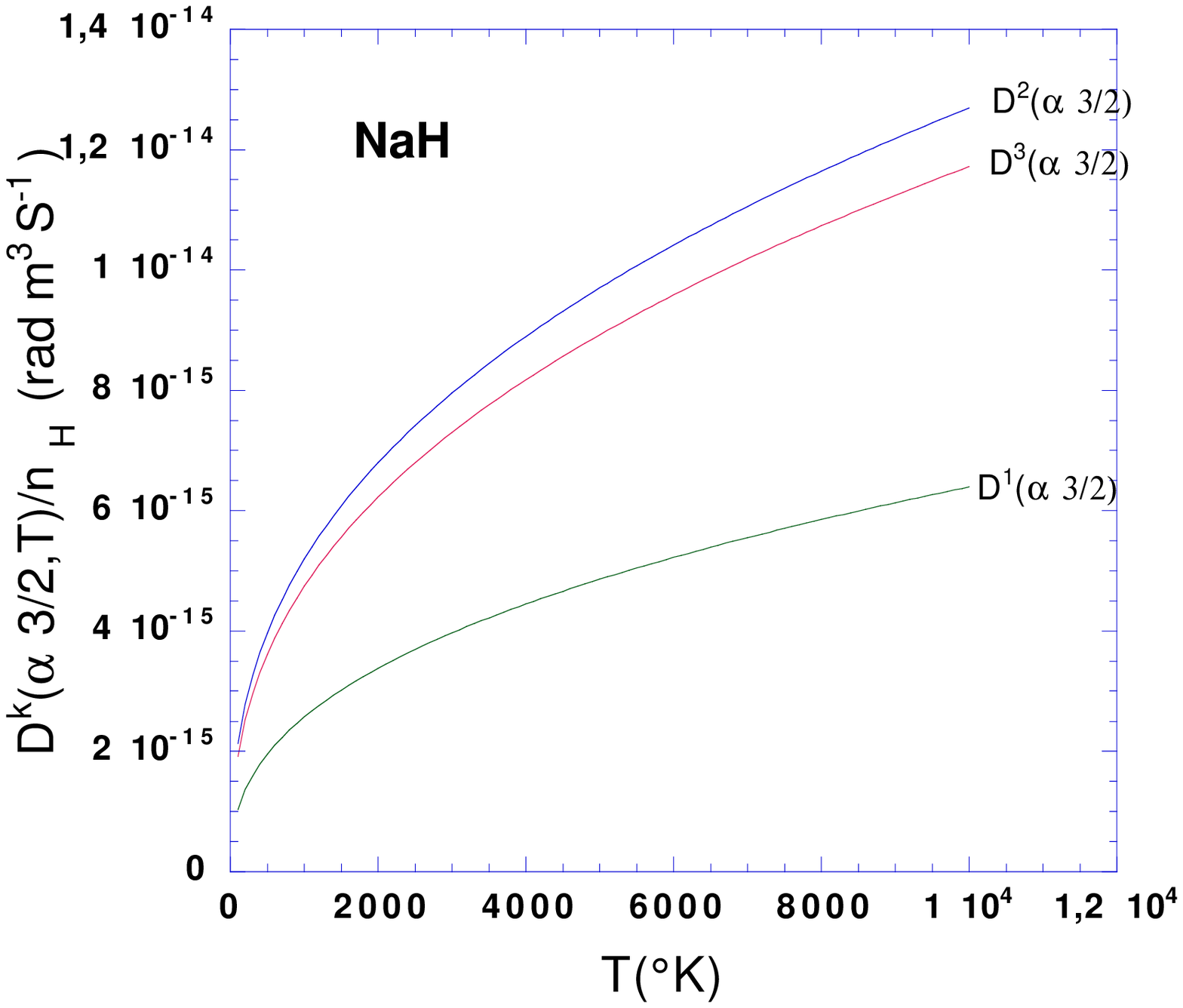}
\end{center}
\caption{Depolarization rates as a function of temperature.}
\label{depolarizationrates}
\end{figure}

Figures \ref{depolarization} and \ref{depolarizationrates} display the  depolarization cross sections (computed directly) and the corresponding depolarization rates for the CaI $4 p\; ^1P_1 (n^* \simeq 2.077)$, MgI $3 
p\; ^1P_1 (n^* \simeq 2.030)$  and the NaI $3 p \; ^2P_{3/2}(n^* \simeq 2.117)
$ states.
\begin{table}
\begin{center}
\begin{tabular}{|l|c|c|c|c|r|}
\hline
$n^*$ & $\lambda^1(\alpha 1)$ & $\lambda^2(\alpha 1)$ & $\lambda^1(\alpha \frac{3}{2})$ & $\lambda^2(\alpha \frac{3}{2})$ & $\lambda^3(\alpha \frac{3}{2})$ \\
\hline
$1.5$ &0.269& 0.256 &0.201& 0.246&	0.240\\
\hline 
$1.6$ &0.270& 0.257 &0.207 & 0.246&	0.234\\
\hline 
$1.7$ &0.265& 0.247 &0.211 & 0.244&	0.235\\
\hline 
$1.8$&0.259& 0.249 &0.213 & 0.242&	0.240\\
\hline
$1.9$&0.256& 0.242 &0.215 & 0.240&	0.234\\
\hline 
2&0.251& 0.235 &0.216 & 0.237&	0.235 \\
\hline
$2.1$ &0.253& 0.237 &0.220 & 0.235&0.218\\
\hline 
$2.2$ &0.248& 0.235 &0.219 & 0.232&0.223\\
\hline 
$2.3$ &0.246& 0.235 &0.221 & 0.231&0.221\\
\hline 
$2.4$ &0.242& 0.231 &0.220 & 0.230&0.225\\
\hline 
$2.5$ &0.240& 0.224 &0.221 & 0.228&0.224\\
\hline 
$2.6$ &0.242& 0.227 &0.223 & 0.228&0.220\\
\hline 
$2.7$ &0.241& 0.223 &0.224 & 0.228&0.226\\
\hline 
$2.8$ &0.244& 0.226 &0.228 & 0.230&0.223\\
\hline 
$2.9$ &0.249& 0.223 &0.221 & 0.230&0.208\\
\hline 
$3$ &0.243& 0.229 &0.229 & 0.2208&0.225 \\ 
\hline  
\end{tabular}
\end{center}
\caption{Velocity exponents $\lambda^k(\alpha J)$ corresponding to the cross sections of Table 2.}
\label{lambda^k}
\end{table}

To obtain the depolarization rates we must integrate over the Maxwell distribution of velocities.
\begin{table}
\begin{center}
\begin{tabular}{|l|c|c|c|c|r||}
\hline
         &\multicolumn{2}{|c|}{T=5000K}&\multicolumn{2}{|c|}{T=6000K}\\
\cline{2-3}\cline{3-5}
$n^*$  & $D^2(\alpha 1)$/$n_H$ & $D^2(\alpha \frac{3}{2})$/$n_H$ & $D^2(\alpha 1)$/$n_H$ & $D^2(\alpha \frac{3}{2})$/$n_H$ \\
\hline
1.5 & 0.5338 & 0.3931&0.5705&0.4206 \\
\hline
1.6 & 0.6135 &0.4477&0.6556&0.4789 \\
\hline
1.7 & 0.7188&0.5220&0.7679&0.5585 \\
\hline
1.8 & 0.8363&0.6047&0.8944&0.6470 \\
\hline
1.9 & 0.9716&0.7000&1.0395&0.7492 \\
\hline
2.  & 1.1260&0.8084&1.2050&0.8653 \\
\hline
2.1  &1.2971&0.9289&1.3885&0.9944 \\
\hline
2.2  &1.4844&1.0599&1.5891&1.1349 \\
\hline
2.3  &1.6875&1.2021&1.8066&1.2872 \\
\hline
2.4  &1.9043&1.3563&2.0390&1.4523 \\
\hline
2.5  &2.1452&1.5270&2.2982 &1.6352 \\
\hline      
2.6  &2.4065&1.7086&2.5774&1.8297 \\
\hline
2.7  &2.6814&1.9010&2.8709&2.0348 \\
\hline
2.8  &2.9634&2.1015&3.1695&2.2469 \\
\hline
2.9  &3.2665&2.3097&3.5001&2.4707 \\
\hline
3.  &3.5842&2.5237&3.8410&2.6992\\
\hline
\end{tabular}
\end{center}

\caption{Depolarization rates for $l=1$. Each column: $S=0$ and $J=1$; $S=\frac{1}{2}$ and $J=\frac{3}{2}$ for T=5000K and T=6000K, as a function of $n^*$. Depolarization rates are given in $10^{-14}$ $\textrm{rad.} \;  \textrm{m}^3 \; \textrm{s}^{-1}$.}
\label{rates50006000}
\end{table} 
Using equation (\ref{eq44}), it can be shown that for a  mean velocity $\bar{v}=\sqrt{\frac{8 k T}{\pi \mu}}$, the depolarization rates can be 
expressed as (Anstee \& O'Mara \cite{Anstee1})

\begin{eqnarray} \label{eq46}
D^k(\alpha J, T)& =&(\frac{4}{\pi})^{(\frac{1}{2}\lambda^k(\alpha J))} \Gamma (2-\frac{1}{2}\lambda^k(\alpha J))v_0  \sigma^k(\alpha J, v_0) \nonumber \\
&& \times \;  (\frac{\bar{v}}{v_0})^{1-\lambda^k(\alpha J)} 
\end{eqnarray} 

Since $\displaystyle f(v)$ depends on the reduced mass $\displaystyle \mu$, the
depolarization rate  is specific to a given atom. In order to obtain  depolarization rates for any level of any atom by simple interpolation we applied the 
approximation: $\displaystyle \mu = m_H, m_H$ being the hydrogen mass.
We verified that this approximation introduces a less than $4 \%$ error in
the depolarization rates.

Table \ref{rates50006000} gives 
the alignment depolarization rates as a function of $n^*$ for the temperatures
of 5000 K and 6000 K.
\subsection{Comparison with other methods}
\begin{figure}[h]
\begin{center}
\includegraphics[width=8 cm]{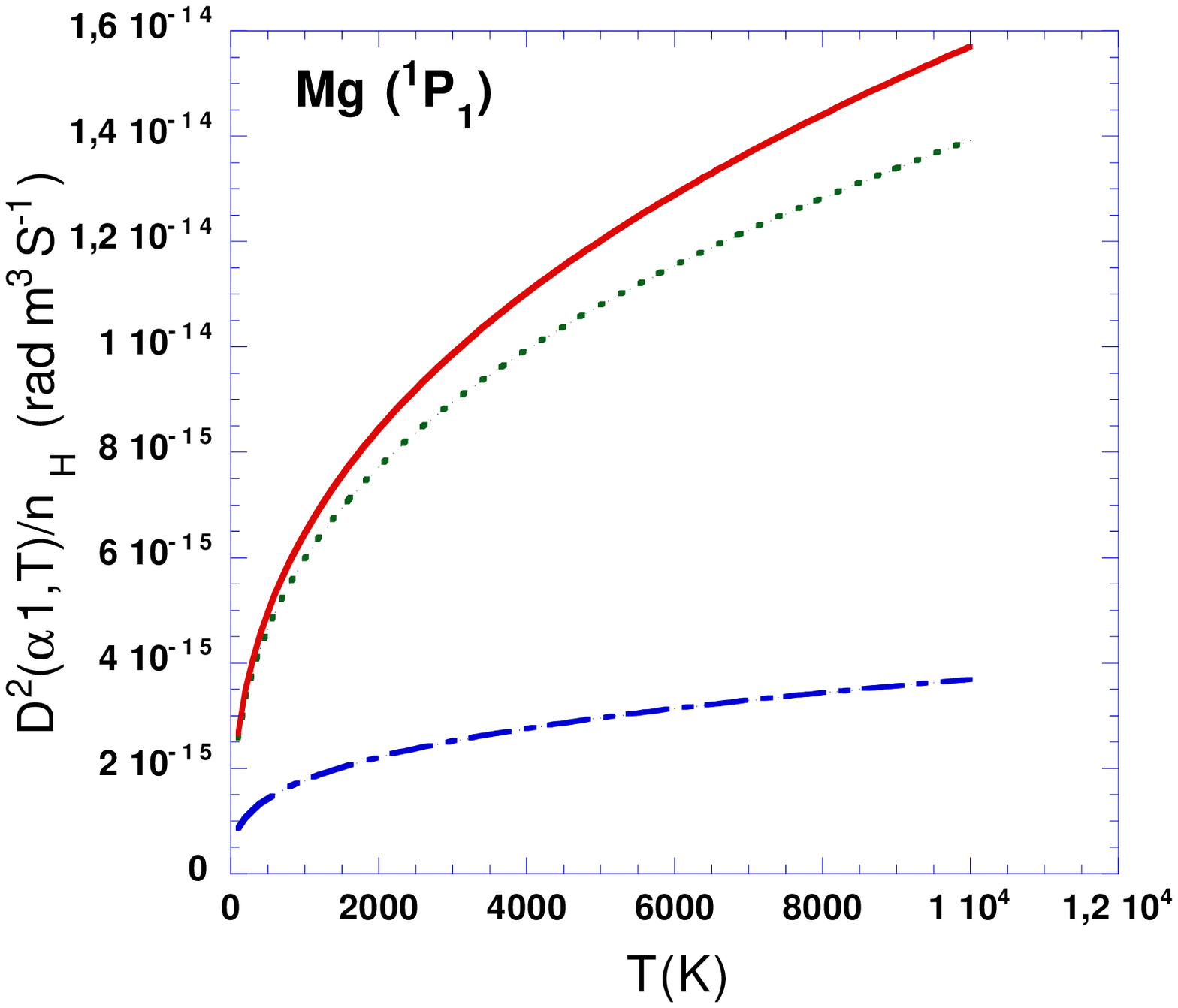} 
\end{center}
\begin{center}
\includegraphics[width=7.5 cm]{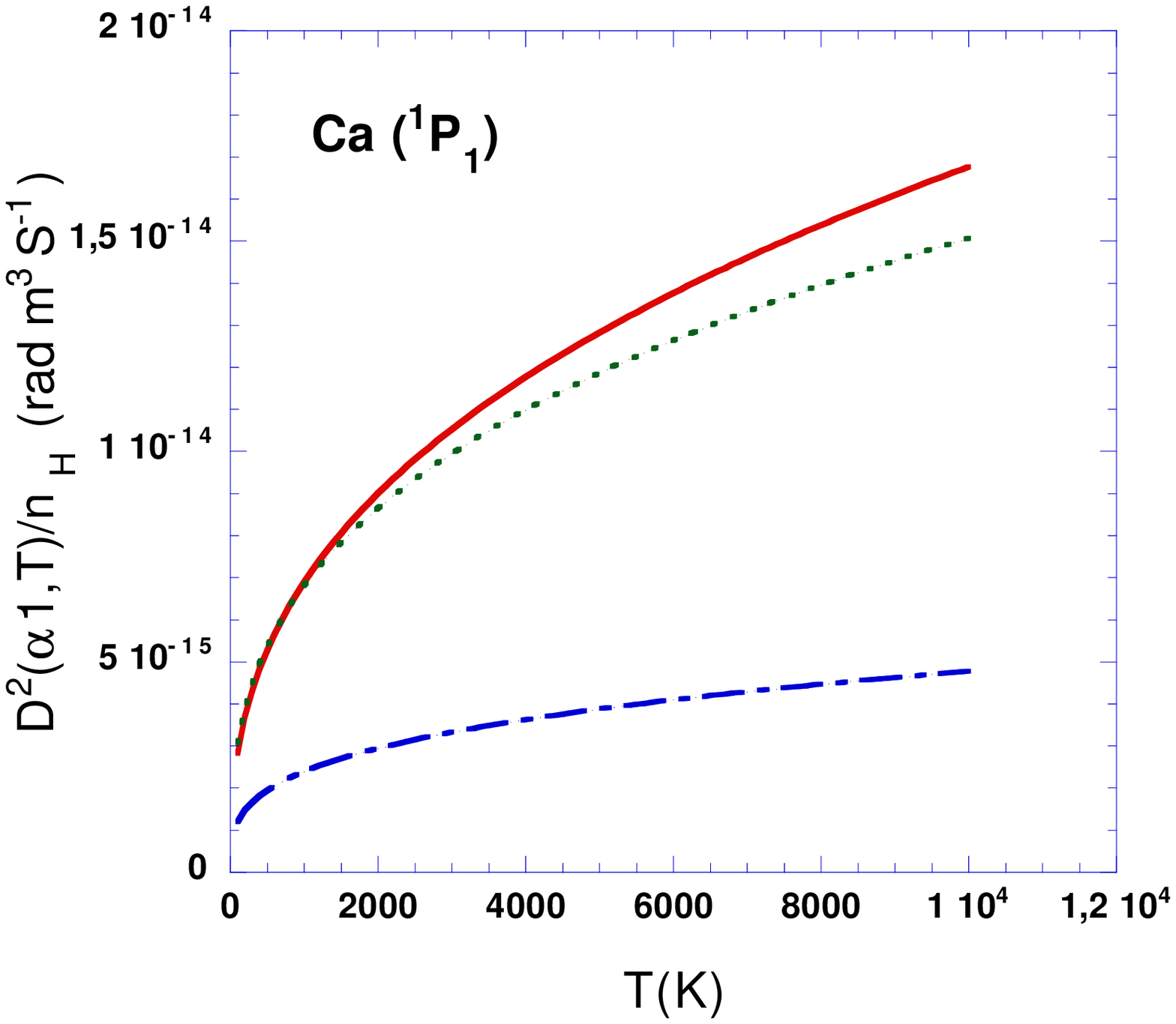}
\end{center}
\begin{center}
\includegraphics[width=8 cm]{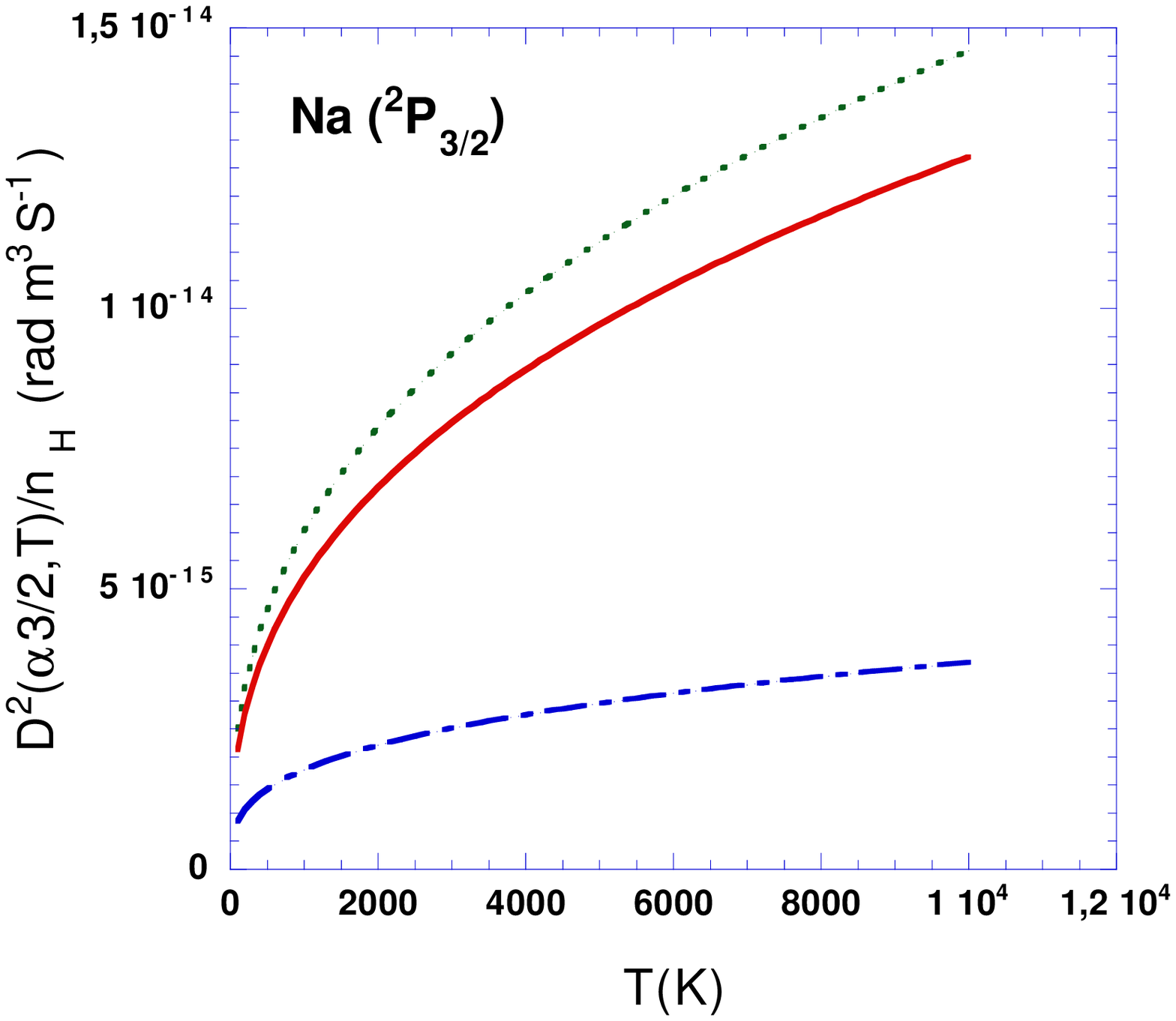}
\end{center}
\caption{Depolarization rates for $k=2$ as a function of temperature. Full lines: our results; dotted lines: quantum chemistry calculations (Kerkeni \cite{Kerkeni2}); dot-dashed lines  Van der Waals approximation.}
\label{depolarizationratesfunctiontemperature}
\end{figure}

In Figure \ref{depolarizationratesfunctiontemperature} we compare our alignment depolarization rates with quantum chemistry depolarization rates (Kerkeni \cite{Kerkeni2}) and the alignment depolarization rates obtained by 
replacing the RSU potential by the Van der Waals potential, $\displaystyle V = -\frac{C_6}{R^6}$, where $\displaystyle C_6$ is the Van der Waals constant.

We  display  only the $k=2$ case which is related to the linear 
depolarization (alignment). Our results show rather good agreement with quantum chemistry calculations.  
However, the Van der Waals potential underestimates the depolarization 
cross section. To explain this we now examine the sensitivity of the 
depolarization cross section  to the RSU potential.  

\clearpage

\section{Dependence of depolarization  cross sections on potentials } \label{sec8}
As a first check on the sensitivity of the results to the nature of the potential 
we introduce a localised perturbation to the interaction potential. We have chosen the case CaI 4$p$ for the check. The interaction potential is multiplied by a  Gaussian magnification factor of the form (Anstee \& O'Mara \cite{Anstee1}) 
\begin{eqnarray} \label{eq47}
G(R)=1+\exp(-(R-R_0)^2).
\end{eqnarray} 

The $\sigma$-symmetry interaction $(1s,4p\sigma)$ is more important than the $\pi$-symmetry interaction $(1s,4p\pi)$. Therefore only the $(1s,4p\sigma)$ interaction has been altered. $R_0$ is 
the position of the maximum perturbation.

\begin{figure}[h]  
\begin{center}
\includegraphics[width=7 cm]{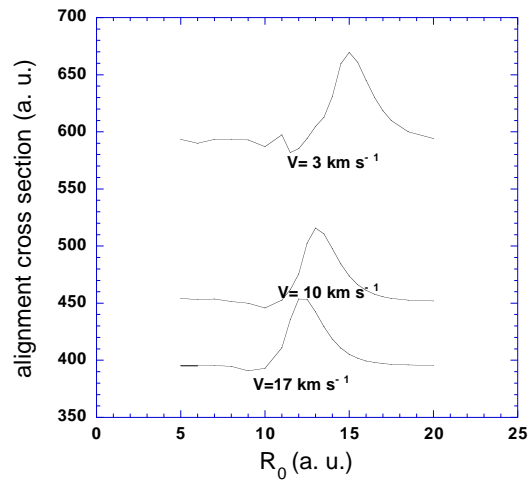}
\end{center}
\caption{Depolarization cross section enhancement due to a Gaussian local 
perturbation of the potential. Cross sections are calculated for  
the relative velocities: $3 \; \textrm{km} \; \textrm{s}^{-1}$, $ 10 \; \textrm{km} \; \textrm{s}^{-1}$ and $ 17 \; \textrm{km} \; \textrm{s}^{-1}$ .}
\label{depolarizationcrosssectionenhancement}
\end{figure}

 Fig. \ref{depolarizationcrosssectionenhancement} shows the effect of the $R_0$ variation on the depolarization cross section for three different velocities. We remark that the enhancement effect on the 
cross section shows a peak in the intermediate range 
($10 a_0\leq R_0\leq 18 a_0$). We conclude that the depolarization cross section depends most strongly on the intermediate region of the potential. It is for this reason that  the Van der Waals interaction potential, which is inaccurate in this intermediate region, underestimates the depolarization cross sections. 

\begin{figure}[h]  
\begin{center}
\includegraphics[width=8 cm]{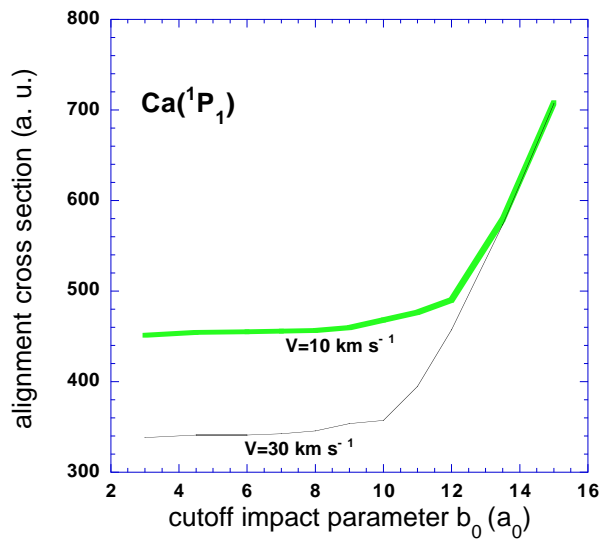}
\end{center}
\caption{Behaviour of the depolarization cross sections from equation $(\ref{eq43})$ 
with the cut-off $b_0$. For a relative velocities  $v=10 \; \textrm{km} \; \textrm{s}^{-1}$ 
and $v=30 \; \textrm{km} \; \textrm{s}^{-1}$.}
\label{Behaviours}
\end{figure}

  To study the sensitivity of the cross section to  close collisions, a 
second check is made. We have calculated the alignment cross sections by varying the cutoff $b_0$ in 
equation (\ref{eq43}). Fig.  \ref{Behaviours} shows the effect of the cutoff variation at $v=10 \; \textrm{km} \; \textrm{s}^{-1}$ and $v=30 \; \textrm{km} \; \textrm{s}^{-1}$. For $b_0 < 8 a_0$ the cutoff variation does not have any 
effect on the cross section calculations. 

Therefore it can be concluded that short-range collisions which involve ionic and exchange 
interactions do not influence the depolarization cross sections and hence the 
depolarization rates. The principal differences between the RSU potentials and those from quantum chemistry occur at  small interatomic separations. Since the
important impact-parameters are at medium interatomic distances, this 
explains why we obtain a rather good agreement between the ABO  
and quantum chemistry results. However, a more detailed study would be desirable
but  is outside the scope of the present paper because the exact quantum 
chemistry potentials are needed for quantitative comparison.

\section{Conclusion}  \label{sec9}
In conclusion, we have shown that the ABO theory is a powerful tool for
quickly calculating a great number of depolarization cross sections for collisions with atomic hydrogen in the physical conditions of the solar atmosphere. The present paper is limited to $p$ states for testing the method. Comparison with  quantum chemistry results  gives us the conviction that it would
 be useful  to extend
our calculations to higher $l$-values which are also of importance for solar 
line polarization. Although less accurate  than quantum 
chemistry calculations, an extension to heavy atoms and 
higher $l$-values is easier and faster within the present semi-classical 
method. We will also extend our 
semi-classical theory to  depolarization cross sections by collisions 
with singly ionized atoms. This will be the subject of  further papers.

\begin{acknowledgements}
This paper is dedicated to Jim O'Mara's memory, who 
followed with interest the progress of the present work and suddenly
passed away on April 27, 2002. The authors  thank the  referee for valuable suggestions which improved the 
presentation of this work. We are indebted to B. Kerkeni who has 
communicated a number of her results before publication.
\end{acknowledgements}

\end{document}